\documentstyle[12pt]{article}
\begin{document}
\vspace{-2.0cm} 
\begin{flushright}
TPI-MINN-96/03 \\
NUC-MINN-96/5-T \\
hep-lat/9603003 \\
\end{flushright}
\bigskip
\begin{center}
\large {\bf A STUDY OF THE BULK PHASE TRANSITIONS OF THE 
SU(2) LATTICE GAUGE THEORY WITH MIXED ACTION}  \\

\bigskip

\large {Rajiv V. Gavai\footnote {On sabbatical leave from the Tata 
Institute of Fundamental Research, Homi Bhabha Road, Mumbai 400005, India. 
E-Mail: gavai@tpi1.hep.umn.edu and gavai@theory.tifr.res.in}} \\
\bigskip
Theoretical Physics Institute \\
School of Physics and Astronomy \\ 
University of Minnesota \\
Minneapolis, MN 55455, U.S.A. \\

\bigskip
\bigskip
{\bf ABSTRACT}\\ \end{center}
\bigskip
\noindent  
Using the finite size scaling theory, we re-examine the nature of the bulk
phase transition in the fundamental-adjoint coupling plane of
the SU(2) lattice gauge theory at $\beta_A = 1.25$ where previous finite
size scaling investigations of the deconfinement phase transition showed
it to be of first order for temporal lattices with four sites.  Our 
simulations on $N^4$ lattices with N=6, 8, 10, 12 and 16 show an absence 
of a first order bulk phase transition.  We find the discontinuity 
in the average plaquette to decrease approximately linearly with $N$. 
Correspondingly, the plaquette susceptibility grows a lot slower with the 
4-volume of the lattice than expected from a first order bulk phase 
transition.

\newpage

\begin{center}
1. \bf INTRODUCTION \\
\end{center}
\bigskip

The lattice regularization of quantum field theories is a gauge invariant 
non-perturbative tool to investigate long distance phenomenon, such
as the confinement of quarks, and to extract the various low energy
properties of quantum chromodynamics, such as the hadronic spectrum.  As
with other regularizations, there is a lot of freedom in defining a
lattice field theory.   In particular, a variety of different choices of
the lattice action correspond to the same quantum field theory in the
continuum.  While most of the numerical simulations are performed for
the Wilson action\cite{Wil} for the gauge theories, other choices, some
motivated by the desire to find a smoother continuum limit, have also
been used.  Indeed, since these actions differ merely by irrelevant
terms in the parlance of the renormalization group, they must give rise to
the same physical results.  In view of the necessity of using finite
lattices and the not-so-small lattice spacings in computer simulations,
investigations with different actions can provide an independent check on
the cut-off independence of the physical results.

From a more theoretical point of view, investigations with different
lattice actions could provide clues in understanding the physics of
confinement.  Since confinement of quarks can be explicitly shown on 
the lattice in the strong coupling region, one naively expects a smooth
passage to the weak coupling regime without any phase transitions in
order for confinement to persist in the continuum limit.
Bhanot and Creutz \cite{BhaCre}, extending the form of the
action proposed by Wilson by adding an adjoint coupling term , showed that 
confinement could survive even though the phase diagram of the mixed
action, shown in Fig. 1,  has the so-called bulk phase transitions along
the solid lines.   The termination of the lower line at a finite adjoint
coupling, $\beta_A$, in Fig. 1 allows a smooth path between the confining and
the asymptotically free phases.  The proximity of this end point to the
$\beta_A = 0$ line, which defines the Wilson action, has commonly been
held responsible for the abrupt change from the strong coupling 
region to the scaling region for the Wilson action.  Its relative
closeness to the $\beta_A=0$ line for the $SU(2)$ theory compared to the
$SU(3)$ theory has been thought of as a possible reason for the
shallower dip in the corresponding non-perturbative $\beta$-function
obtained\cite{subet} by Monte Carlo Renormalization Group methods.
In view of the fact that rather small lattices were used to obtain the
phase diagram in Fig. 1, it therefore appears necessary to re-examine the 
phase diagram on bigger lattices and with better statistics.  The work
reported in this manuscript is a step in that direction.

Another major motivation for re-examining the phase diagram comes
from our work\cite{us1,us2,us3} on the same mixed action at non-zero
temperatures. Along the $\beta_A =0.0$ axis, several finite temperature 
investigations have shown the presence of a second order deconfinement 
phase transition. Its critical exponents have been 
shown\cite{EngFin} to be in very good agreement with those of the three 
dimensional Ising model.  Effective field theory arguments for the order
parameter were used by Svetitsky and Yaffe\cite{SveYaf} to conjecture
the finite temperature SU(2) gauge theory and the three dimensional
Ising model to be in the same universality class.  The verification of
this universality conjecture thus strengthened our analytical understanding 
of the deconfinement phase transition.  However, following the deconfinement 
phase transition into the extended coupling plane by simulating the extended
action at finite temperature, we surprisingly found that:  

\begin{enumerate} 

\item[{a]}] The deconfinement transition was of second order, and in agreement 
with the universality conjectured exponents, only up to 
$\beta_A \approx 1.0$. It became definitely of first order for large 
enough $\beta_A (\ge 1.25)$. 

\item[{b]}] There was no evidence of an another separate bulk transition at
larger $\beta_A$, as suggested by Fig. 1. 

\end{enumerate}

Using asymmetric lattices, $N_\sigma^3 \times N_\beta$, with $N_\beta =
2$-8 and $N_\sigma = 8$-16, we obtained the following key results which
lead us to the conclusions above:
 
\begin{enumerate}

\item[{a]}] Only one transition was found on all the lattices studied. 
The deconfinement order parameter acquired nonzero large value at the 
transition and showed a clear co-existence of both phases at the transition 
point for larger $\beta_A$.

\item[{b]}] The same critical exponent which established the transition
to be in the Ising model universality class for $0.0 \le \beta_A \le 1.0$ 
became equal to the space dimensionality (=3), as a first order deconfinement
phase transition would have, for larger $\beta_A$.

\end{enumerate}

As argued in Ref. \cite{us1} already, 
the apparent coincidence of the trajectory of the
deconfinement phase transition for $N_\beta=4$ with the lower arm of the
bulk phase diagram of Fig. 1 itself suggests a possible explanation of 
these results: The first order bulk transition overshadows the second
order deconfinement phase transition.  It is therefore mandatory to confirm
the existence of the first order bulk transition on bigger 
symmetric lattices and with better statistics than that of Ref.
\cite{BhaCre}.  In this paper, we undertake this task by simulating the
mixed action on symmetric $N^4$ lattices, with $N = 6$-16.
The organization of the paper is as follows: In section 2  we 
define the action we investigate and briefly recapitulate the definitions 
of various observables used and their scaling laws. We present the detailed
results of our simulations in the next section and the last section  contains
a brief summary of our results and their discussion.

\bigskip
\begin{center}
2.  THE MODEL AND THE OBSERVABLES \\
\end{center}
\bigskip

The lattice action is constrained only by a) the gauge
invariance and b) the limit of zero lattice spacing which must coincide
with the continuum form of the action.  Infinitely many different forms
satisfying these criteria can be written down.  Bhanot and Creutz extended the
Wilson action to a form described by the action,

\begin{eqnarray}
S = \sum_P \left( \beta \left(1 - {1\over2} Tr_F U_P \right) + 
         \beta_A \left(1 - {1\over3} Tr_A U_P \right) \right)~~~~~.
\label{ea}
\end{eqnarray}

Here $U_P$ denotes the directed product of the basic link
variables which describe the gauge fields, $U_\mu(x)$, around an
elementary plaquette $P$. $F$ and $A$ denote that the respective traces are
evaluated in the fundamental and adjoint representations respectively.  
We use the formula $Tr_A U = |Tr_F U | ^2 -1$.

Comparing the naive classical continuum limit of eq. (\ref{ea}) 
with the standard $SU(2)$ Yang-Mills action, one obtains
\begin{eqnarray}
{1 \over g^2_u} = {\beta \over 4} + {2\beta_A \over 3}~~.~~
\label{gu} 
\end{eqnarray}
Here $g_u$ is the bare coupling constant of the continuum theory.  
Since the asymptotic scaling equation for the mixed action can be easily
written down in terms of $g_u$ with a $\Lambda$-parameter that depends
on the ratio of $\beta$ and $\beta_A$, it is clear that the introduction 
of a non-zero $\beta_A$ does not affect the continuum limit:
the theory for  each $\beta_A$, including the usual Wilson theory for
$\beta_A = 0.0$ flows to the same critical fixed point, $g^c_u = 0$,
in the continuum limit and has the same scaling behavior near the
critical point.  

As mentioned already in the introduction, Bhanot and Creutz\cite{BhaCre} 
found that the lattice theory defined by the extended action of eq.(\ref{ea}) 
has a rich phase structure (Fig. 1). Along the $\beta = 0$ 
axis, it describes the $SO(3)$ model which has a first order phase transition 
at $\beta_A^{crit} \sim 2.5$.  At $\beta_A = \infty$, it describes the $Z_2$ 
lattice gauge theory again with a first order phase transition at 
$\beta^{crit} = {1\over2} \ell n(1 + \sqrt{2})$ $\approx$ 0.44 \cite {Weg}. 
Ref. \cite {BhaCre} found that these first order transitions extend into the 
($\beta$,$\beta_A$) plane, ending at an apparent 
critical point located at (1.5,0.9).  These transition lines are shown in 
Fig. 1 by continuous lines. The qualitative aspects of this phase diagram 
were also later reproduced by mean field theory \cite {AlbFly} 
and large N\cite{OgiHor} and strong coupling\cite{DasHel} expansions.

We simulated the mixed action above on $N^4$ lattices, with $N =6$, 8,
10, 12, and 16.  Periodic boundary conditions were used in all the four
directions.  The partition function is, as usual, given by,
\begin{eqnarray}
Z = \int \prod_{x, \mu} dU_\mu(x)~~ \exp ( -S) ~.~
\label{parfun}
\end{eqnarray}
We used the simple Metropolis algorithm and tuned it to have an acceptance
rate $\sim 30$-40 \%.  The expectation values of the observables were 
recorded every 20 iterations to reduce the autocorrelations.  Errors
were determined by correcting for the autocorrelations and also by
binning. The observables monitored were the average plaquette, P,
defined as the average of $Tr_F U_P$/2 over all independent plaquettes,
and the average of $L(\vec n)$ over the three dimensional lattice
spanned by $\vec n$, where $L$ is defined by
 \begin{eqnarray}
 L(\vec n) = {1\over2} Tr_F \prod^{N}_{\tau=1} U_0 (\vec n,\tau)~.~
 \label{pol}
 \end{eqnarray}
Here $U_0 (\vec n,\tau)$ is the timelike link at the lattice site 
$(\vec n,\tau)$ and due to the symmetry of our lattices any direction could 
be identified as the time direction.  One sees that the $L$ corresponds to
the usual order parameter\cite{McSv} for the deconfinement transition on
a lattice with $N$ temporal sites but with also $N$ spatial sites.  We
will comment later on the utility of such an observable on symmetric
lattices.  In order to monitor the nature of the bulk phase transition,
we also define the plaquette susceptibility: 
\begin{eqnarray}
\chi_N = 6 N^4 (\langle P^2 \rangle - \langle P \rangle^2)~.~
\label{chi}
\end{eqnarray}

According to the finite size scaling theory\cite{Barb}, the peak of the 
plaquette susceptibility at the location of the bulk transition we wish 
to study should grow on an $N^d$ lattice like
\begin{equation}
\chi^{max}_{N} \propto N^{\omega}~.~
\label{chifs}
\end{equation}
For a second order transition, $\omega=\alpha/\nu$, where $\alpha$ and $\nu$ 
characterize the growth of the plaquette susceptibility and the correlation 
length near the critical coupling (temperature) on an infinite lattice.
If the phase transition were to be of first order instead, then
one expects the exponent $\omega = d = 4$, corresponding to the dimensionality
of the space \cite{ChLaBi}.  In addition, of course, the average plaquette is
expected to exhibit a sharp, or even discontinuous, jump and the corresponding
probability distribution should show a double peak structure in case of
a first order phase transition.

\bigskip 

\begin{center}
3.  \bf RESULTS OF THE SIMULATIONS\\
\end{center}
\bigskip
 
Our Monte Carlo simulations were done using the Metropolis algorithm
on $N^4$ lattices with $N$= 6, 8, 10, 12 and 16. The many different values of 
$N$ were chosen to study the finite size scaling behavior of the 
theory and to compute the critical exponent $\omega$.  For verifying the
nature of the lower arm of the phase diagram in Fig. 1, any value of
$\beta_A$ between 0.9 and 2.0 would be suitable.  Considering, however,
the results of Ref. \cite{us3}, where the deconfinement transition at 
$\beta_A = 1.25$ was shown to be of first order using $N^3 \times 4$ 
lattices with $N = 8$, 10, and 12, we chose $\beta_A = 1.25$ although
we also attempted additional simulations at $\beta_A = 1.5$.
Histogramming techniques\cite{FerSwe} were used to extrapolate to 
nearby $\beta$ values for estimating the height and location of the peak 
of the plaquette susceptibility.  

Figs. 2 and 3 display the distributions of the average plaquette $P$
at $\beta_A=1.25$ on $6^4$, $8^4$, $10^4$ and $12^4$, $16^4$ lattices 
respectively.  The values of the fundamental coupling $\beta$ at which
these runs were made are 1.2147, 1.2179, 1.2182, 1.2183 and 1.2184
respectively and the corresponding number of measurements, each
separated by 20 iterations, are 135000, 107000, 109000, 32500 and 13250.
Thus Fig. 2 is based on equally high statistics runs.  While one
observes a double peak structure on each of the lattices used, it is
clear that the distance between the peaks, i.e., the discontinuity in
the plaquette, $\Delta P$, decreases with increasing lattice size.
Furthermore, one can also conclude from Fig. 2 that the valley between
the peaks becomes shallower with the increase in lattice size.  Both
these observations are, of course, precisely opposite of what one
expects for a first order bulk phase transition.  
The results in Fig. 3  for the bigger lattices are also in accord with 
both these trends established in Fig. 2, although they are based on
rather modest statistics.
Indeed, we have actually displayed the results for two neighboring
couplings $\beta = 1.2183$ and 1.2184 on the $16^4$ lattice to highlight
any possible double peak structures in these runs.  If one disregarded
the smaller peaks in these runs as statistically marginal, then the
actual discontinuity in $\Delta P$ on the $16^4$ is most likely smaller
than that suggested by Fig. 3.   In Fig. 4, we show the corresponding
plaquette histograms for the finite temperature case of Ref. \cite{us3}
where a first order deconfinement phase transition was established at the {\it
same} $\beta_A = 1.25$. This was based on the determination of the
critical exponent for the Polyakov loop ($L$) susceptibility which was
found to be equal to the space dimensionality, three, in
that case.  Note that the relative increase in (spatial) volume in going
from the smallest to the biggest lattice in Fig. 4 is comparable to that
for the two smaller lattices in Fig. 2.  However, one sees that the
peaks in Fig. 4 hardly move and moreover, the valley structure deepens as
the spatial volume is increased. Both these observations are
in full accord with the expectations
for a first order phase transition.  Of course, the key difference in
these figures is that the temporal extent is kept fixed for Fig. 4
whereas it too is increased along with the spatial extent in Figs. 2 
and 3.  Thus the deconfinement phase transition does exhibit the behavior
expected of a first order phase transition while the bulk phase
transition at the same $\beta_A$ does not.

The above qualitative observation of a lack of a bulk first order-like
behavior can be made quantitatively more firm by using conventional
ideas and methods.   Using smooth curves to approximate the peaks in
Figs. 2-4, one can estimate the location of each peak and deduce the size
of the discontinuity, $\Delta P$, in the average plaquette in each case.  
These results are given in Table 1 and are also shown in Fig. 5 as a 
function of $1/N$, where the simulations were done on an $N^4$ lattice.  
Also listed in Table 1 are the results for the finite temperature case 
of Fig. 4. The errors reflect the bin sizes in Figs. 2-4.  While the
$\Delta P$ in the finite temperature case is constant, the data 
on symmetric $N^4$ lattices are consistent with a linear fall off 
with $1/N$ and seem to suggest a zero discontinuity in the average
plaquette on a finite but large lattice, of the $O(70^4)$.  
A linear extrapolation to infinite lattice predicts a 
$\Delta P ( \infty) = -0.011 \pm 0.006$, which too suggests a lack of
any transition although a second or higher order transition will be
difficult to rule out. Fig. 6
exhibits the plaquette susceptibility as a function of $\beta$ for the
$6^4$, $8^4$ and $10^4$ lattices. The data point in each case
corresponds to the runs shown in Fig. 2 and the error bars reflect the
increase in the autocorrelation length with lattice size.
One can see the that our long runs on each of the lattices are
indeed very close to the location of the peak. Thus minimal systematic errors 
are expected from the histogramming extrapolation in the location
and the heights of the peaks listed in Table 2.  Noting that the increase 
in the 4-volume, $N^4$, is respectively a factor of 3.16 and 7.72 compared
to the smallest lattice, it is clear that the increase in plaquette
susceptibility is far short of that needed for a first order bulk
transition.  A fit of the peak heights to eq. (\ref{chifs}) yields an
$\omega = 2.09 \pm 0.31$ which is to be contrasted with the space
dimensionality, 4.  The fitted value  of $\omega$
is consistent with a linear decrease in the 
discontinuity $\Delta P$ seen in Fig. 5, as can be seen from eq. (\ref{chi}).  
One can use this exponent to predict the peak heights for the $N=12$ and 16
lattices.  As can be seen from Table 2, these values, 89 and 163 respectively, 
compare favorably with the Monte Carlo results. A better determination
of these peak heights is computationally very hard due to both the large
lattice sizes and the increase in autocorrelations.
Nevertheless, it seems clear that the
bulk transition at $\beta_A = 1.25$, if there is one, is not a first
order transition.  At the very least, this suggests the endpoint of the
lower arm in Fig. 1 to
be at $(\beta, \beta_A) = (1.2184, 1.25)$, although a still higher
$\beta_A$ and correspondingly smaller $\beta$ seems more likely. 
In view of the results shown in Fig. 4, and discussed in more
details in Ref. \cite{us3}, a first order deconfinement phase transition
does seem to exist at $\beta_A = 1.25$, indicating that the
deconfinement transition (for $N_\beta = 4$ but infinite spatial volume)
turns first order before $\beta_A$ is large enough for a first order
bulk phase transition to exist.  One is therefore lead to conclude that
the change in the order of the deconfinement phase transition is indeed
a real finite temperature effect at large $\beta_A$.

We have also studied the histograms of the average Polyakov loop, $L$
defined above, to look for a possible deconfinement phase transition on
these lattices and at $\beta_A = 1.25$ .  One expects large 
corrections due to finite volume since $VT^3$ is unity in stead of 
being very large compared to one.  An expected consequence thus is 
significantly wider distributions for $L$, making the critical coupling,
where the distribution develops a lot flatter peak or a multi-peak 
structure, shift.  We found that the histograms for $L$ were peaked at
zero for all the $\beta$ values discussed above.  Increasing $\beta$ a
little, the peak flattened and developed a three peak structure which
was however not as sharp as the plaquette histograms.  As an example,
let us quote that such a determination lead to $\beta_c(N=6) \simeq 1.2179$,
which should be compared with the $\beta_c$ obtained from the peak of
the plaquette susceptibility and given in Table 2.  One may be tempted
to interpret this as a sign that the deconfinement transition is
splitting away from the bulk transition and moving to larger $\beta$.  On
the other hand, it is not uncommon for different observables and thus
different definitions to yield different estimates for the location of the
same transition on a {\it finite} lattice.  Only in the thermodynamic limit
is it necessary for all the estimates to coincide.  As argued already, such
differences are all the more natural since $VT^3 =1 $.  It therefore
appears to us that even on these symmetric lattices either the deconfinement
phase transition is the only transition or at least it is coincident
with a (second or higher order) bulk transition signaled by the 
rapid changes in the average action.   
It should be noted that a {\it decrease} in the plaquette
discontinuity, as seen in Fig. 5,  would be naively consistent with the
first option since an increase in $N$ for a constant critical
temperature would amount to decreasing the lattice 
spacing in that case and a decrease in plaquette discontinuity could then
be a way to hold the corresponding latent heat constant.  In that case,
the bulk finite size scaling arguments, involving a scaling with the
4-volume $N^4$ would not apply.  In stead, it would be necessary to check 
the scaling by holding $N_\beta$ fixed and with a large $VT^3$.
Precisely such an exercise was carried out in Ref. \cite{us3} and it did
establish a first order deconfinement phase transition.

We have also attempted to repeat the above exercise at a larger 
$\beta_A = 1.5$ to find out whether a first order bulk phase transition
exists there.  As the size of discontinuity on the smaller lattices,
$6^4$ and $8^4$, also increases with the increase in $\beta_A$, it
became difficult to perform any meaningful finite size scaling analysis.
In particular, we found that the ordered and random starts at $\beta =
1.045$ on these lattices remained separate even after 50000
measurements, corresponding to 1 million sweeps.  This calls for a use of
better algorithms which will encourage tunnelings between the two
states and thus permit a reliable finite size scaling study. The
plaquette discontinuities from our runs were found to be 0.22352 (15)
and 0.22291 (12), showing a miniscule decrease of 0.00061 (19).   While
this is an encouraging sign for the existence of a first order bulk
phase transition, the results of Fig. 7 show an amusing correlation of
the deconfinement phase transition with it.  Fig. 7 shows the hysteresis
effects in the average plaquette $P$ and the Polyakov loop $L$ on $N^4$
lattices with $N = 4$, 6, 8, 10 and 12 at $\beta_A = 1.5$.  Starting from
a $\beta = 0.95$ with a disordered start for the gauge variables, 2000
iterations were performed at each $\beta$ at an interval of $\delta
\beta =0.01$ to thermalize and then the observables were recorded over
the next 8000 iterations.  Similarly a run was begun from $\beta=1.15$
with an ordered start to obtain the other branch.  A good agreement between
the two curves outside the metastable area is an indication that the
metastabilities are real.   In spite of the large fluctuations in $L$ for 
larger $\beta$, it appears that both $L$ and $P$ on all the lattices 
undergo strong changes at about the same $\beta$.  The size of plaquette
discontinuity suggested by Fig. 7 is approximately the same as mentioned
above and it seems to remain unchanged even on a $12^4$ lattice.   It
should perhaps be noted that the decrease in the discontinuity in $L$ is
related to the well known observation of the decrease of $L$ in the high
temperature phase with $N$ ( or temporal lattice size).  Unfortunately,
the big metastable region in Fig. 7, related to the large discontinuity
in $P$, makes it very hard to ascertain whether one is dealing with two
transitions here or one and what finite size scaling properties they
have (or it has).

\bigskip
\begin{center}
4.  \bf SUMMARY AND DISCUSSION \\
\end{center}
\bigskip

The phase diagram of the mixed action of eq. (\ref{ea}) in the
fundamental and adjoint couplings, $\beta$ and $\beta_A$, has been
a crucial input in understanding many properties of the $SU(2)$ and
$SU(3)$ lattice theories and their continuum limit.  The cross-over to 
the scaling region from the strong coupling region, as well as the dip 
in the non-perturbative $\beta$-function have been attributed to the 
location of the end point of the line of bulk first order phase transition.  
In fact, even the relative shallowness of the dip for the $SU(2)$ case
compared to the $SU(3)$ case is thought to be due to the closeness of
the corresponding end point to the $\beta_A = 0$ Wilson axis.
The recently observed change of the order of the deconfinement phase 
transition for the $SU(2)$ lattice gauge theory for lattices with four 
(and two) temporal sites could also be due to the seemingly puzzling
coincidence of the bulk first order line with the deconfinement line for
large enough $\beta_A$.

The phase diagram of Fig.1, taken from Ref. \cite{BhaCre}, was obtained
on smaller lattices, $5^4$-$7^4$, and with modest statistics. 
We simulated the extended action on $N^4$ lattices with $N$ = 6, 8, 10,
12, and 16 at $\beta_A=1.25$ and 1.5.
The choice of these adjoint couplings was based on the results of Ref.
\cite{us1,us3} where the deconfinement transition was shown to be of
first order for $N_\beta$ = 4 using finite size scaling theory.  In
particular, the susceptibility for the order parameter, the Polyakov
loop, was shown to increase linearly with spatial volume at
$\beta_A$=1.25; it grew approximately as the two-third power of the
spatial volume for small $\beta_A$ which is similar to the Ising model
in three dimensions which has a second order phase transition.
We found that the plaquette distributions do exhibit a double peak
structure on the symmetric lattices as well.  However, the major
difference was that the peaks appear to approach each other and the
intervening valley appears to become shallower as the lattice size $N$
is increased at the {\it same} $\beta_A=1.25$.  Quantitatively, this was
reflected in a much slower increase in the plaquette susceptibility with
the 4-volume than would be expected for a first order bulk phase
transition and a linear decrease in the size of the discontinuity of the
plaquette, $\Delta P$, with $N$. The critical exponent $\omega$, defined
in eq. (\ref{chifs}), was found to be $2.09 \pm 0.31$ in contrast to the
expected value of 4 for a first order bulk phase transition.  While 
a linear extrapolation of our data on $\Delta P$ suggests that 
it vanishes already on finite, $O(70^4)$ lattices, we are unable to rule
out a second or higher order bulk phase transition at $\beta_A=1.25$. We
conclude that the endpoint of the bulk line in Fig. 1 is most likely at
a $\beta_A \ge 1.25$ with a correspondingly smaller $\beta = 1.2184$
or lesser.  It also seems therefore that the change of the order of the
deconfinement phase transition for $N_\beta = 4 $ at $\beta_A \simeq 1.25$ is
unlikely to be influenced by any (first order) bulk phase transition.  
As argued in Ref. \cite{us3}, strong coupling arguments do suggest 
precisely this, namely, the increase in $\beta_A$ changes the effective 
potential for the order parameter to allow a first order deconfinement 
phase transition at sufficiently large $\beta_A$ and thus
the bulk dynamics need not play any role in such a change.

Our simulations at the larger $\beta_A = 1.5$ were inadequate to test in
a similar manner using finite size scaling theory whether the transition
there is a first order bulk phase transition or not.  The large size of
the plaquette discontinuity on the smaller lattices meant that the
Metropolis algorithm was inefficient in effectively sampling both the
states, thus aborting our attempts to check whether $\omega =4$. On the
other hand, the decrease in $\Delta P$ with $N$ appeared to be much
smaller than at $\beta_A=1.25$.  Hysteresis runs on a variety of lattice
sizes showed that the deconfinement order parameter too jumps at the
transition rather abruptly, suggesting that the deconfinement transition
is either very close to the bulk transition or even coincident.  It
would be very interesting to decipher the finite size scaling behavior
at these larger $\beta_A$ to distinguish the two very different
transitions.  Indeed, even in the case of the $SU(3)$ gauge theory,
where simulations\cite{Blum} with the mixed action have yielded a
separation of the line of the deconfinement phase transitions from the
line of bulk phase transitions with increasing $N_\beta$, a convincing
demonstration of such a separation would really come from similar finite
size scaling investigations.

\bigskip 

\begin{center}
6.  \bf ACKNOWLEDGMENTS \\
\end{center}
\bigskip 

The computations reported here were performed on the DEC Alpha
machines of the Tata Institute of Fundamental Research, Mumbai and the
Theoretical Physics Institute, University of Minnesota, Minneapolis
and the CRAY-2 of the Minnesota Supercomputer Institute, Minneapolis. I
would like to thank the staff at these institutes for their support. I
gratefully acknowledge the hospitality I received in
TPI, Minneapolis, especially from Profs. J. Kapusta and L. McLerran.
This work was supported by the U. S. Department of Energy under the
grant DE-FG02-87ER40328.

\newpage 
\pagestyle{empty}
\begin{table}
\begin{center}
{Table 1}
\end{center}
The average values of the plaquette discontinuity $\Delta P$ \\ 
at $\beta_A = 1.25$ on symmetric $N^4$ lattices and asymmetric \\
$ N^3 \times 4$ lattices.  The data for latter are taken \\
from Ref. \cite{us3} \\ \\
\medskip
\begin{tabular}{|c|c|}
\hline
                 &                             \\
~~~~~~~~~Lattice Size~~~~~~~~~~      &~~~~~~~~~~~ $\Delta P$~~~~~~~~~   \\
                 &                              \\
\hline \hline
                 &                              \\
$6^4$            &           0.111(4)     \\
                 &                         \\
\hline 
                 &                         \\
$8^4$            &            0.081(4)     \\
                 &                         \\
\hline
                 &                         \\
$10^4$           &            0.060(4)     \\
                 &                         \\
\hline
                 &                         \\
$12^4$           &            0.0495(40)   \\  
                 &                         \\
\hline
                 &                          \\
$16^4$           &            0.036(4)     \\  
                 &                          \\
\hline\hline
                 &                          \\
$8^3 \times 4$   &            0.102(4)      \\  
                 &                          \\
\hline       
                 &                          \\
$10^3 \times 4$  &            0.099(4)     \\  
                 &                          \\
\hline       
                 &                          \\
$12^3 \times 4$  &            0.096(4)     \\  
                 &                          \\
\hline       
\end{tabular} 
\end{table}

\newpage 
\pagestyle{empty}

\begin{table}
\begin{center}
{Table 2}
\end{center}
The values of $\beta$ at which simulations were performed \\
on $N^4$ lattices at $\beta_A = 1.25$ , $\beta^{crit.}$ and the \\
height of the plaquette susceptibility peak, $\chi^{max}_N$.\\ \\
\medskip
\begin{tabular}{|c|c|c|c|}
\hline
            &            &                       &       \\
Lattice Size     &$~~~~~~~\beta~~~~~~~$ & $~~~~~~~~\beta_c~~~~~ ~$ & $~~~~~~~~~\chi^{max}_N~~~$ \\
            &              &                       &       \\
\hline \hline
      &    &       &          \\
     $6^4$  & 1.2147  & 1.21497 &  20.73(1.14) \\
\hline
      &    &       &          \\
     $8^4$  & 1.2179  & 1.21783 &  41.78(4.90) \\
\hline
      &    &       &          \\
     $10^4$  & 1.2182  & 1.21826 &  51.70(11.42) \\
\hline
      &    &       &          \\
     $12^4$  & 1.2183  & 1.2183  &  $\sim$ 74    \\
\hline
      &    &       &          \\
     $16^4$  & 1.2184  & 1.21836 &  $\sim$134    \\
\hline
\end{tabular} 
\end{table}
\newpage

\pagestyle{empty}
\begin{center}
\bf{FIGURE CAPTIONS}
\end{center}

\bigskip

\noindent Fig. 1 
The phase diagram of the extended SU(2) lattice gauge theory.  
Taken from Ref. \cite{BhaCre}. 
\bigskip

\noindent Fig. 2
 Probability distribution of the average Plaquette $P$ at
$\beta_A=1.25$ on $N^4$ lattices with $N$=6, 8, and 10 
The values of $\beta$ for these runs are given in the text.
\bigskip

\noindent Fig. 3
Same as in Fig. 2 but for $N$ = 12 and 16.  
\bigskip
 
\noindent Fig. 4
Same as in Fig. 2 but for the asymmetric lattices $N^3 \times 4$ with
$N$ = 8, 10, and 12.
lattices.
\bigskip

\noindent Fig. 5
The discontinuity in the average Plaquette, $\Delta P$, on $N^4$
lattices as a function of $1/N$ for $\beta_A=1.25$. The line denotes a
simple linear fit.
\bigskip

\noindent Fig. 6
Plaquette susceptibility as a function of $\beta$ on $N^4$ lattices with
$N$=6, 8, and 10 for $\beta_A=1.25$.

\bigskip

\noindent Fig. 7
The hysteresis in the average plaquette $P$ and the order parameter for
deconfinement, $L$, as a function of $\beta$.  All the results were
obtained on $N^4$ lattices for $N$=8, 10, and 12 and at $\beta_A =
1.25$.
\bigskip
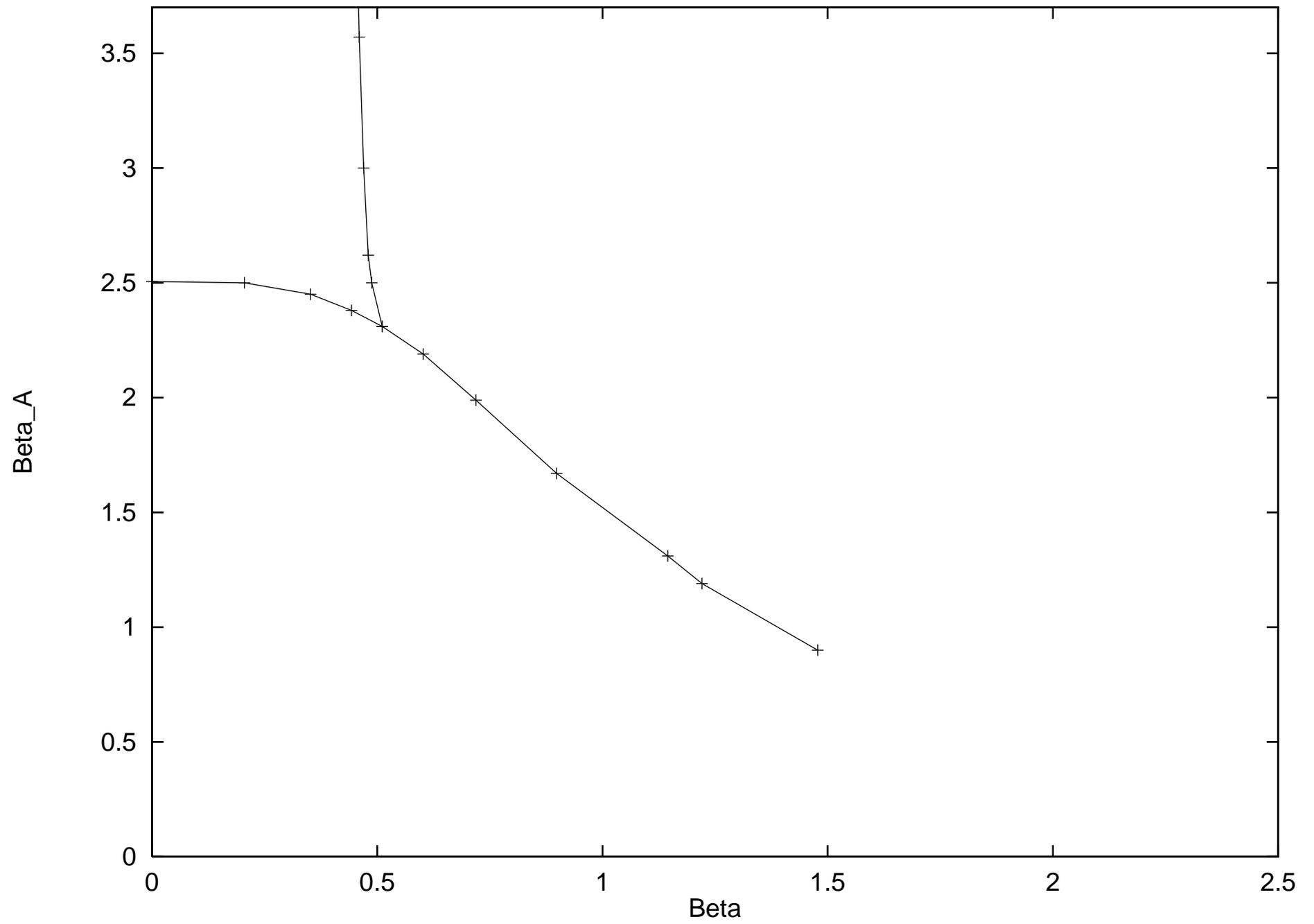
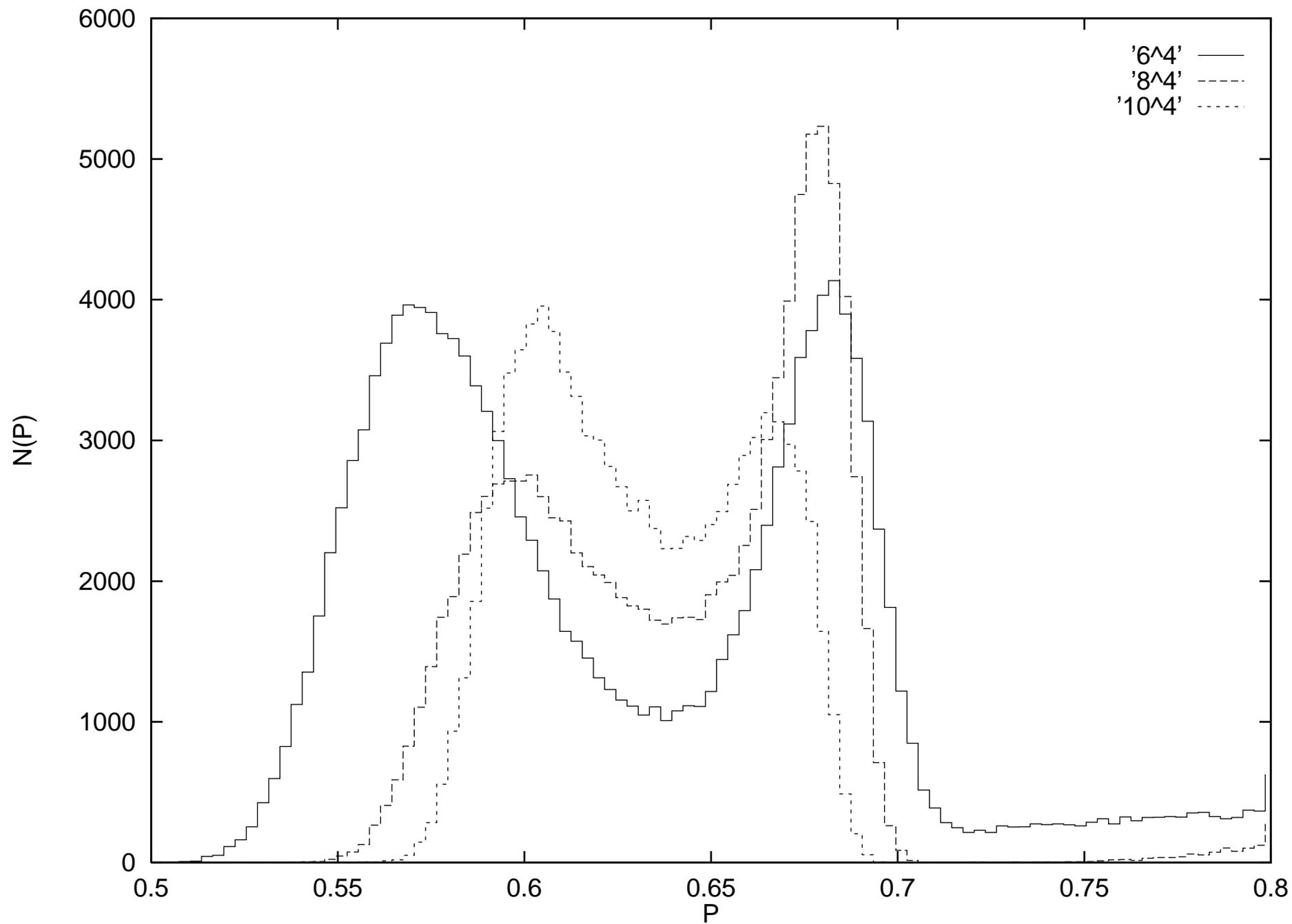
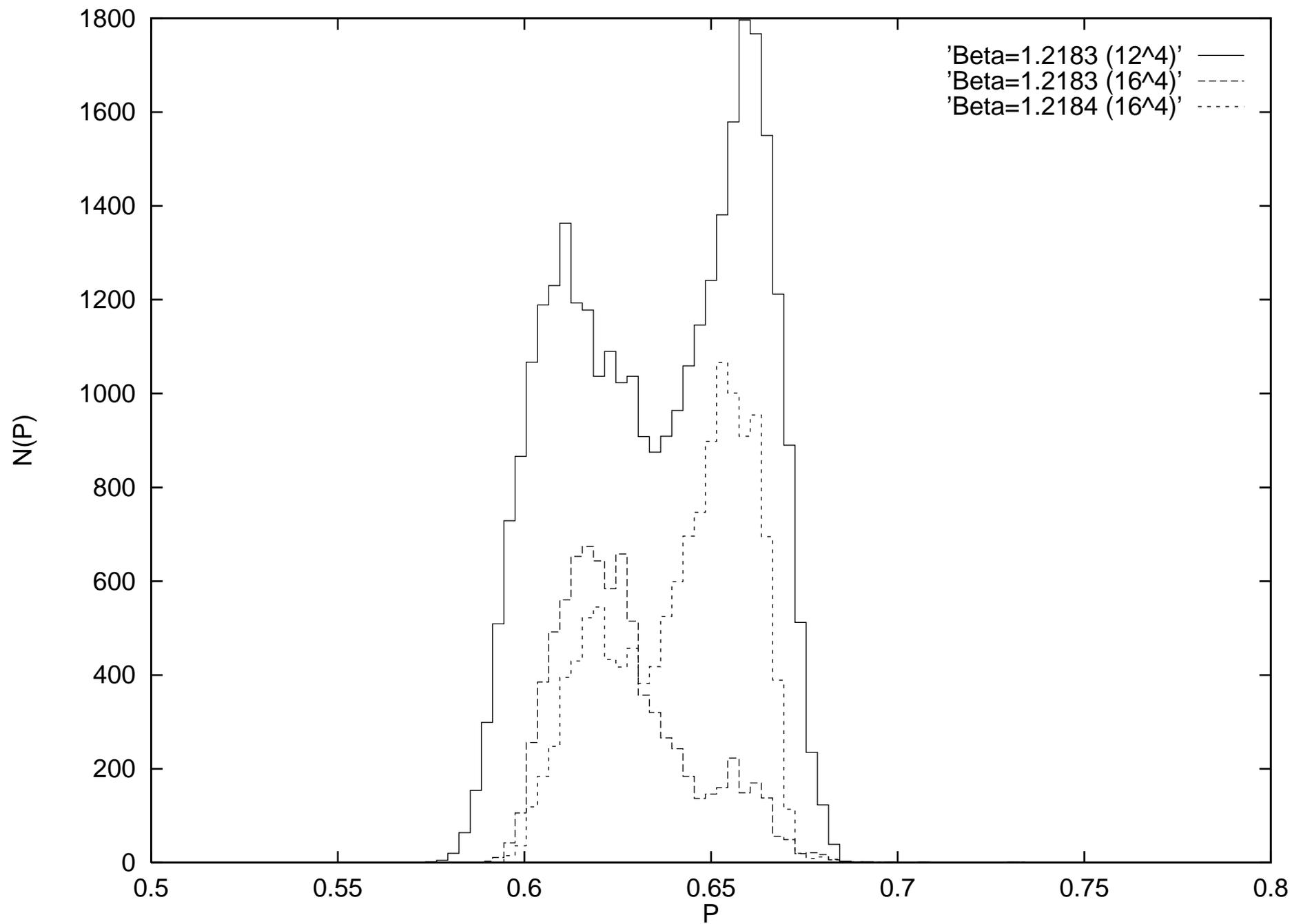
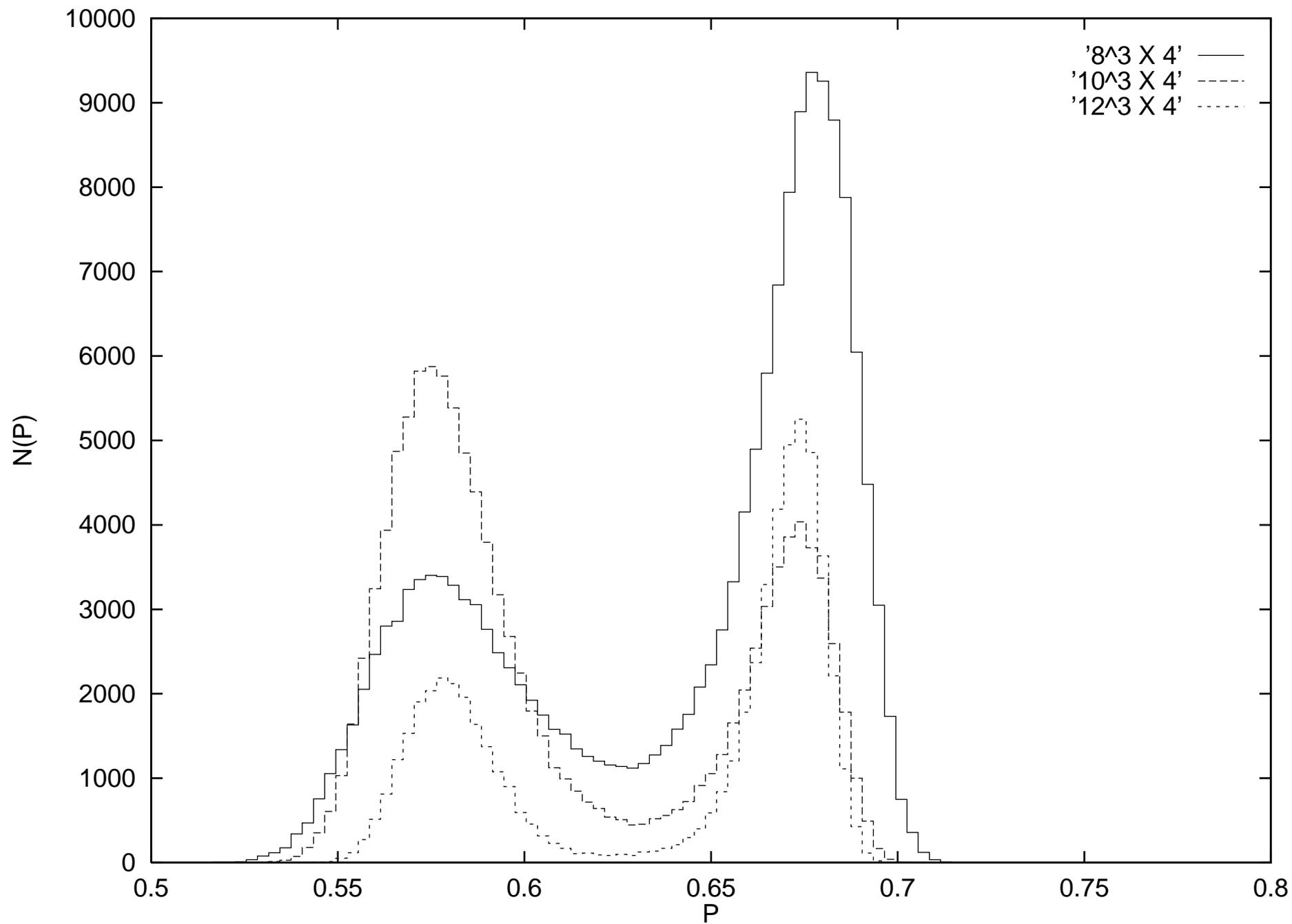
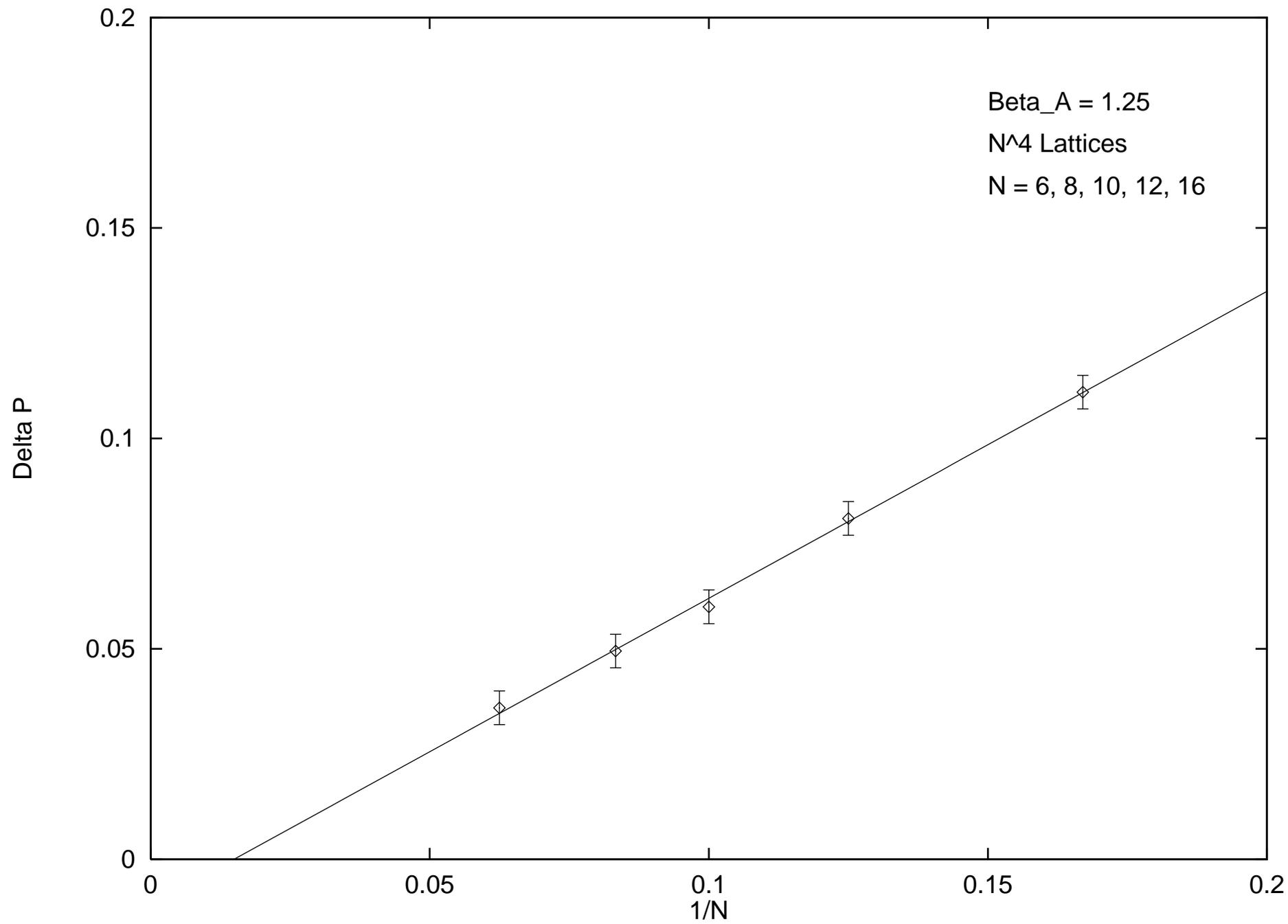
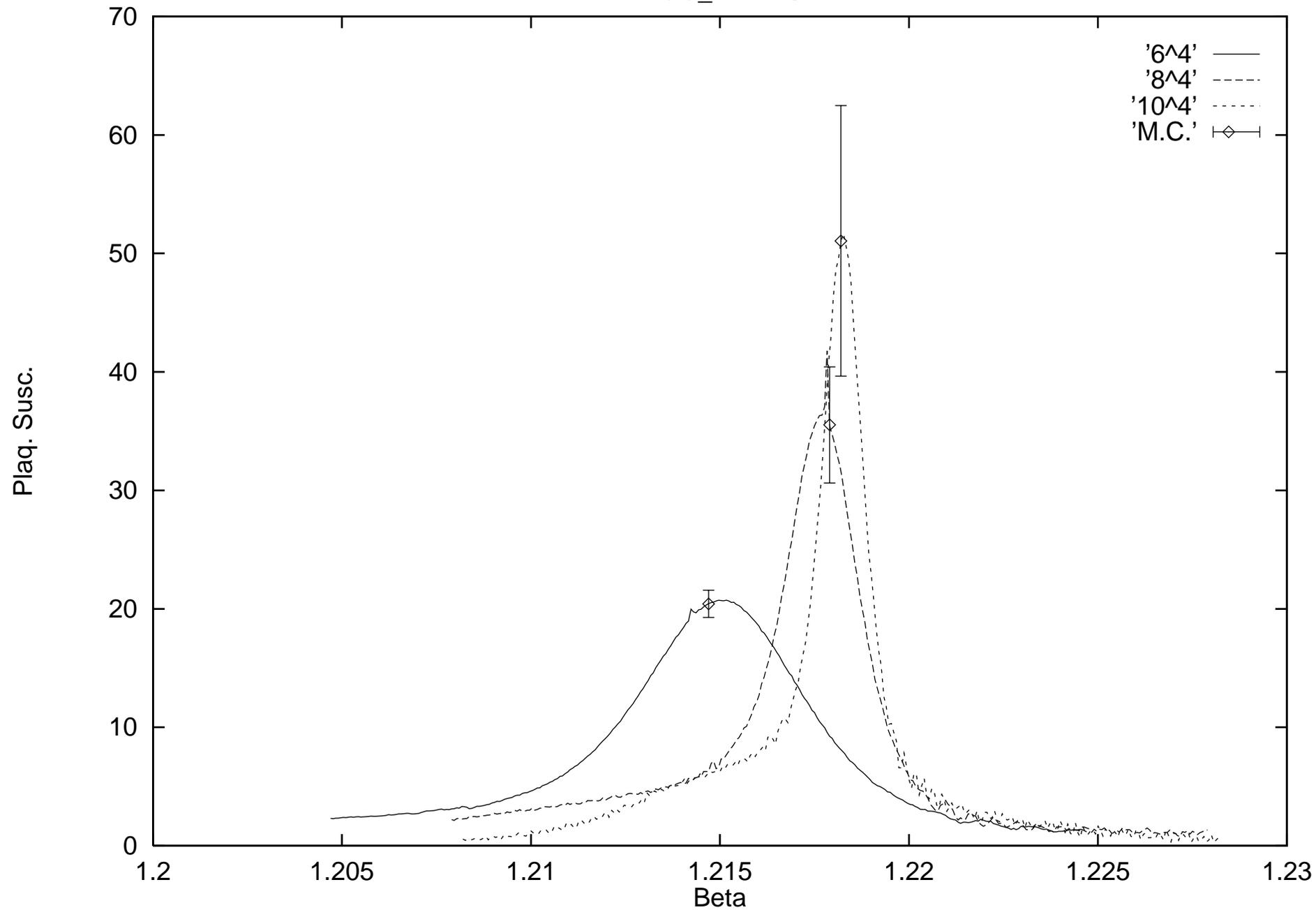
\end{document}